\documentclass[onecolumn]{emulateapj}
\usepackage{amsmath} 
\usepackage{graphicx}
\usepackage{natbib}

\newcommand{\q}{\mathbf{q}}
\newcommand{\x}{\mathbf{x}}
\newcommand{\kk}{\mathbf{k}}
\newcommand{\rr}{\mathbf{r}}
\newcommand{\nablab}{\mathbf{\nabla}}

\begin{document}

\title{Nonlinear Behavior of Baryon Acoustic Oscillations from the Zel'dovich Approximation Using a Non-Fourier Perturbation Approach}
\author{Nuala McCullagh and Alexander S. Szalay}
\affil{The Johns Hopkins University \\ Baltimore, MD 21218}

\begin{abstract}
Baryon acoustic oscillations are an excellent technique to constrain the properties of dark energy in the Universe. In order to accurately characterize the dark energy equation of state, we must understand the effects of both the nonlinearities and redshift space distortions on the location and shape of the acoustic peak. In this paper, we consider these effects using the Zel'dovich approximation and a novel approach to 2nd order perturbation theory. The second order term of the Zel'dovich power spectrum is built from convolutions of the linear power spectrum with polynomial kernels in Fourier space, suggesting that the corresponding term of the the Zel'dovich correlation function can be written as a sum of quadratic products of a broader class of correlation functions, expressed through simple spherical Bessel transforms of the linear power spectrum. We show how to systematically perform such a computation. We explicitly prove that our result is the Fourier transform of the Zel'dovich power spectrum, and compare our expressions to numerical simulations. Finally, we highlight the advantages of writing the nonlinear expansion in configuration space, as this calculation is easily extended to redshift space, and the higher order terms are mathematically simpler than their Fourier counterparts.
\end{abstract}


\section{Introduction}
\label{sec:intro}

Baryon acoustic oscillations (BAO) arise from photon pressure opposing the gravitational collapse of baryonic density perturbations in the baryon-photon plasma before the epoch of recombination \citep{peeblesyu1970,silk1968, sz1970, bondefstathiou1984}. We see the signature of this effect in both the temperature power spectrum of the cosmic microwave background (CMB) \citep{debernardis2000,hanany2000}, and the galaxy power spectrum. The BAO feature was first detected as a peak in the correlation function of the Luminous Red Galaxies (LRGs) in the Sloan Digital Sky Survey (SDSS) by \citet{eisenstein2005}, and was subsequently measured in several other galaxy surveys \citep{cole2005, padmanabhan2007, percival2007, percival2010}. 

The BAO signal is useful because at low redshift, the scale of interest is still largely in the linear regime. The location of the peak thus acts as a ``standard ruler'' with which the expansion history of the universe can be measured as a function of time. This allows us to constrain the equation of state of Dark Energy \citep{seoeisenstein2003}.

In order to accurately constrain the properties of dark energy using BAO, we must understand the effects of nonlinearities and redshift space distortions on the acoustic peak. Linear theory predicts that the amplitude of the peak will scale as the square of the linear growth function, $D(t)$. However, it is important to understand to what extent the nonlinearities contribute to the location and shape of the peak at low redshift. It is also well known that there are first order effects on the correlation function due to the transformation from real to redshift space \citep{kaiser1987, tian2011, blake2011, seo2012}. Understanding how this transformation affects higher order terms will be increasingly important as observations become more precise.

Much work has been done to characterize the nonlinearities using cosmological perturbation theory \citep{jainbertschinger, bouchet1995,bernardeau2002,croccescoccimarro2006, croccescoccimarro2008, matsubaraRPT2008}, and to extend this work into redshift space \citep{matsubaraPTRSD2008, taruya2010, valageas2011}. However, the correlation function has many advantages over the power spectrum when analyzing the BAO signal. Here, we present a method of calculating the first nonlinear term in the real-space correlation function using the Zel'dovich approximation. In Section~\ref{sec:first}, we show the derivation of the nonlinear term from the Zel'dovich approximation. In Section~\ref{sec:second} we show how this is the exact Fourier transform of the nonlinear Zel'dovich power spectrum. We directly compare our calculation to the expression for the nonlinear Zel'dovich power spectrum, originally calculated in \citet{grinsteinwise1987}, and more recently in \citet{valageas2011}. We show in Section~\ref{sec:discussion} that our configuration-space term is in agreement with numerical simulations of the Zel'dovich approximation. We conclude in Section~\ref{sec:conclusion}.

\section{Computation of the nonlinear Zel'dovich correlation function in real space}
 \label{sec:first}

The Zel'dovich approximation maps particles' initial Lagrangian coordinate, $\q$, to their co-moving Eulerian coordinate, $\x$, through the gradient of the linear potential, $\phi(\q)$, and the growth function $D(t)$ \citep{zeldovich1970, shandarin1989}.  The linear potential is related to the linear density field through the Poisson equation:
\begin{align}
\x(\q, t) &= \q-D(t) \nablab_q \phi(\q)\ ,\label{eq:Zel'dovich}\\
 D(t)\nabla_q^2\phi(\q)&=\delta_{L}(\q,t)\ .\notag
\end{align}
In the rest of this paper, we will shorten $D(t)$ to $D$ for convenience. The real-space density is then related to the Jacobian of the transformation from Lagrangian to Eulerian coordinates. 
\begin{align}
\frac{\rho(\x, t)}{\bar \rho}&=\left | \frac{\partial x_i}{\partial q_j} \right |^{-1}=\frac{1}{J(\q, t)} =  1 + \delta(\q(\x))\ ,
\label{eq:euleriandensity}
\end{align}
where $\x$ and $\q$ are related by the Zel'dovich formula (\ref{eq:Zel'dovich}). Here, $\delta$ (without subscript $L$) is the {\em weakly nonlinear} over-density. Equation (\ref{eq:euleriandensity}) for the Eulerian density is only strictly valid before shell-crossing, where the mapping from $\q$ to $\x$ is one-to-one. As is discussed in \citet{kofman1994}, when there is multi-streaming, multiple values of $\q=\q_i$ map to the same point $\x$, thus the Eulerian density at a given point is a sum over all of the streams at that point.
\begin{align}
\rho(\x)&=\sum_{i=1}^{ N_{\text{streams}(\x)}} \varrho(\q_i(\x))\ .
\end{align}
Here, $\rho(\x)$ is the full Eulerian density, and $\varrho(\q_i)$ are the ``single-stream'' densities. However, shell-crossing does not contribute greatly at the large scales we are concerned with, and so we assume the ``single-stream'' density is the full density. We will show that this assumption leads to a result that agrees with the usual perturbative results.

The Jacobian can be written in terms of invariants, $I_1$, $I_2$, and $I_3$, of the symmetric matrix $d_{ij}(\q)$ \citep{zeldovich1970}.
\begin{align}
d_{ij}(\q)&=\frac{\partial^2 \phi(\q)}{\partial q_i \partial q_j}\ ,\\
J(\q, t)&=1-D I_1(\q)+D^2 I_2(\q)-D^3 I_3(\q)\ .
\end{align}
The invariants can be written in terms of the eigenvalues of $d_{ij}(\q)$: $\lambda_1$, $\lambda_2$, and $\lambda_3$.
\begin{align}
I_1(\q)&=\lambda_1+\lambda_2+\lambda_3\ ,\notag\\
I_2(\q)&=\lambda_1\lambda_2+\lambda_1\lambda_3+\lambda_2\lambda_3\ ,\\
I_3(\q)&=\lambda_1\lambda_2\lambda_3\ .\notag
\end{align}
Therefore, the Eulerian overdensity can be expressed by the Taylor expansion of the inverse Jacobian, to any order:
\begin{align}
	\delta(\q,t)&=DI_1(\q)+D^2\left(I_1(\q)^2-I_2(\q)\right)+D^3\left(I_1(\q)^3-2I_1(\q)I_2(\q)+I_3(\q)\right)+...
	\label{eq:lagrangedensity}
\end{align}
Note that the first term in this expansion corresponds to linear theory: $\delta_L=DI_1(\q)$.

However, as our goal in this paper is to calculate the density correlation function at a fixed Eulerian distance, we need to be careful about the details of the difference between the Eulerian and Lagrangian coordinates. Instead of the usual forward relation, writing $\x$ in terms of $\q$, we will need the reverse and express $\q$ with $\x$, using the Taylor expansion around $\x$ in a recursive fashion. 
\begin{align}
	\q(\x)&=\x+D\,\nablab_q \phi(\q(\x))\ .
\end{align}
To zeroth order, $\nablab_q \phi(\q(\x))=\nablab_q \phi(\x)$, but when considering the higher order terms in the density, it is necessary to express $\nablab_q \phi(\q(\x))$ to linear order in $D$.
\begin{align}
	\frac{\partial \phi(\x(\q))} {\partial q_i}&=
	\bigg(\frac{\partial \phi(\q)} {\partial q_i}
	+D\, \sum_{j}\frac{\partial^2 \phi(\q)}{\partial q_i\partial q_j}
	\frac{\partial \phi(\q)}{\partial q_j}\bigg) 
	\Bigg|_{\q=\x}\ .
\end{align}

Because the correlation function, $\xi(\rr)$, is a function of Eulerian distance, $\rr = \x_1-\x_2$, we must express the over-density as a function of the Eulerian coordinate, $\x$.  By expanding the function $\delta(\q)$ about the point $\q = \x$, we arrive at an expression for $\delta(\x)$ that is a power series in $D$. 
\begin{align}
\delta(\x,t)&=\delta\left(\x+D\nablab_q \phi(\q(\x))\right)\ .\label{eq:taylordensity}
\end{align}
A Taylor expansion of the right hand side of Equation (\ref{eq:taylordensity}) gives:
\begin{align}
\delta(\x,t)&=\bigg(\delta(\q,t) +D \sum_i \frac{\partial\phi(\q(\x))}{\partial q_i}\frac{\partial\delta(\q,t)}{\partial q_i}+
 \frac{1}{2}D^2 \sum_{i,j}\frac{\partial^2\delta(\q,t)}{\partial q_i \partial q_j}\frac{\partial \phi(\q)}{\partial q_i}\frac{\partial \phi(\q)}{\partial q_j}\bigg)\Bigg|_{\q=\x}\ .
 \end{align}

 With this, we can write the Eulerian density in terms of the Lagrangian density to third order in $D$:
\begin{align}
	\delta(\x,t)&=\bigg(\delta(\q,t) + D \sum_i \frac{\partial\phi(\q)}{\partial q_i}\frac{\partial\delta(\q,t)}{\partial q_i}+
	D^2\sum_{i,j}\frac{\partial^2 \phi(\q)}{\partial q_i\partial q_j}\frac{\partial \phi(\q)}{\partial q_j}\frac{\partial \delta(\q,t)}{\partial q_i}\notag\\
	&\qquad+ \frac{1}{2}D^2 \sum_{i,j}\frac{\partial^2\delta(\q,t)}{\partial q_i \partial q_j}\frac{\partial \phi(\q)}{\partial q_i}\frac{\partial \phi(\q)}{\partial q_j}\bigg)\Bigg|_{\q=\x}\ .
	\label{eq:eulerdensity}
\end{align}
We use Equation (\ref{eq:lagrangedensity}) to express $\delta$ in terms of the linear quantities $\delta_L$ and $\phi$, where $\delta_L$ is already first order in $D$.

The real-space correlation function in co-moving Eulerian coordinates is:
\begin{align}
\xi(\x_1-\x_2, t)&=\langle \delta(\x_1, t)\delta(\x_2, t)\rangle\ .
\end{align}
This can be written in powers of $D$ using the above expansion of the Eulerian over-density. Because the over-density field is assumed to be a zero-mean Gaussian random field, the odd moments vanish. The correlation function to second order is then:
\begin{align*}
\xi(\rr, t)&=\xi^{(1)}(\rr)D^2+\xi^{(2)}(\rr)D^4+...
\end{align*}
We define the functions:
\begin{align}
\xi_n^m(r)&=\frac{1}{2\pi^2} \int_0^{\infty} \!P_L(k)j_n(kr)k^{m+2}\mathrm{d}k\label{xinm}\text{ ,}
\end{align}
where $j_n$ is the spherical Bessel function of order $n$ and $P_L(k)$ is the linear power spectrum. 
Using this definition, the linear term is:
\begin{align}
\xi^{(1)}(\rr)&=\xi_0^0(r)\text{ ,}
\end{align}
the spherically symmetric Fourier transform of the linear power spectrum. 

$\xi^{(2)}(\rr)$ is the expectation of a sum of products of four terms expressed with the linear quantities. Since these are all Gaussian, the only irreducible terms are second order.  Thus we only need to calculate expectations of the type $\langle a(\q_1) b(\q_2) \rangle$. Mathematica was used to express the various derivatives of $\phi(\q)$ in terms of spherical harmonics, and to calculate the expectation values between them. We illustrate this process in the following example.

\begin{align}
\frac{\partial \phi(\q)}{\partial q_z}&=-\int \!\frac{\mathrm{d}^3k}{(2\pi)^3}\frac{i k_z}{k^2}\hat\delta_L(\kk)e^{i \kk \cdot \q}=-i\sqrt{\frac{4\pi}{3}}\int \!\frac{\mathrm{d}^3k}{(2\pi)^3}\frac{Y_{1}^0(\theta, \varphi)}{k}\hat \delta_L(\kk)e^{i \kk \cdot \q}\ ,\\
\frac{\partial^2 \phi(\q)}{\partial q_z^2}&=\int \!\frac{\mathrm{d}^3k}{(2\pi)^3}\frac{k_z^2}{k^2}\hat\delta_L(\kk)e^{i \kk \cdot \q}=\frac{\sqrt{4\pi}}{3}\int\!\frac{\mathrm{d}^3k}{(2\pi)^3} \left(Y_0^{0}(\theta, \varphi)+\frac{2}{\sqrt{5}}Y_2^0(\theta, \varphi)\right)\hat \delta_L(\kk)e^{i \kk \cdot \q}\ .
\end{align}
\begin{align}
\left \langle \frac{\partial \phi(\q_1)}{\partial q_z}\frac{\partial^2 \phi(\q_2)}{\partial q_z^2}\right\rangle&=-\frac{4\pi i}{3^{3/2}(2\pi)^6}\iint\! \mathrm{d}^3k_1 \mathrm{d}^3k_2 e^{i (\kk_1\cdot \q_1 + \kk_2 \cdot \q_2)} \frac{\langle \hat \delta_L(\kk_1)\hat \delta_L(\kk_2)\rangle}{k}\ , \notag \\
&\qquad \times Y_{1}^0(\theta_1, \varphi_1)\left(Y_0^{0}(\theta_2, \varphi_2)+\frac{2}{\sqrt{5}}Y_2^0(\theta_2, \varphi_2)\right)\ .
\end{align}
\begin{align}
\langle \hat\delta_L(\kk_1)\hat\delta_L(\kk_2)\rangle&=(2\pi)^3\delta_D(\kk_1+\kk_2)P_L(k_1)\ .
\end{align}
So we must calculate the integrals
\begin{align}
\int \!Y_l^m(\theta, \varphi)Y_{l'}^{*m'}(\theta, \varphi)e^{i \kk \cdot (\q_1-\q_2)}\mathrm{d}\Omega_k \ .
\end{align}
We use the plane-wave (Rayleigh) expansion for $e^{i \kk \cdot (\q_1-\q_2)}$ to compute these integrals in terms of spherical Bessel functions, $j_n(kr)$, where $\rr=\q_1-\q_2$. In general, the integral is zero for $m\neq m'$.
\begin{align}
\left \langle \frac{\partial \phi(\q_1)}{\partial q_z}\frac{\partial^2 \phi(\q_2)}{\partial q_z^2}\right\rangle&=-\frac{1}{5(2\pi^2)}\int_0^{\infty} \!P_L(k) \left (3 j_1(kr)\mathcal{P}_1(\mu)\notag-2 j_3(kr)\mathcal{P}_3(\mu)\right) k\,\mathrm{d}k\ ,\\
&=-\frac{3}{5}\mathcal{P}_1(\mu)\xi_1^{-1}(r)+\frac{2}{5}\mathcal{P}_3(\mu)\xi_3^{-1}(r)\ .
\end{align}
where $\mu$ is the cosine of the angle between $\hat r$ and $\hat z$, and $\mathcal{P}_l(\mu)$ are the Legendre polynomials. In the expansion of the over-density, Equation (\ref{eq:eulerdensity}), each term is evaluated at $\q=\x$, so the expectation values in the correlation function become functions of the Eulerian distance, $r=|\x_1-\x_2|$.

In the full calculation, there are many seemingly serendipitous cancellations of terms, which in the end give us a relatively simple expression. As we found out, any small mistake in the Mathematica script can lead to a far more complicated expression with the wrong asymptotics, so it is important to understand the expected behavior of the full correlation function to check the result. Firstly, the expectation value of the over-density should vanish to second order. Secondly, the correlation function should be isotropic in real space. Finally, we consider the asymptotic behavior of the expression -- at zero lag, it must be finite, as this is the variance of the over-density field, and on large scales, the expression must tend to zero. Requiring this behavior rules out many erroneous results, and was a useful debugging tool.

The full expression for the nonlinear Zel'dovich correlation function in Eulerian coordinates is:
\begin{align}
	\xi^{(2)}(\rr)&=-\frac{1}{3}\xi_0^{-2}(0)\xi_0^2(r)+ \frac{19}{15}\xi_0^0(r)^2+\frac{34}{21}\xi_2^0(r)^2
	+\frac{4}{35}\xi_4^0(r)^2 - \frac{16}{5}\xi_1^{-1}(r)\xi_1^1(r)\notag\\
	&\qquad-\frac{4}{5}\xi_3^{-1}(r)\xi_3^1(r)+\frac{1}{3}\xi_0^{-2}(r)\xi_0^2(r)
	+\frac{2}{3}\xi_2^{-2}(r)\xi_2^2(r) \ .
\label{eq:nlcf}
\end{align}
At zero lag, this takes the value:
\begin{align}
\xi^{(2)}(0)&=\frac{19}{15}\xi_0^0(0)^2=\frac{19}{15}\sigma_0^4\ .
\end{align}
where $\sigma_0^2$ is the variance of the linear over-density field. As expected, the nonlinear term is finite at $r=0$. As $r\rightarrow \infty$, each set of terms within the expression tends to zero, due to the behavior of the spherical Bessel functions, so the full expression tends to zero as expected.  Note that this expression contains products of the functions $\xi_n^m(r)$.  In the next section, we show that this is the exact Fourier transform of the nonlinear Zel'dovich power spectrum. 

\section{Nonlinear Power Spectrum}
\label{sec:second}

The normalization we use for the power spectrum requires that we rewrite the nonlinear Zel'dovich power spectrum given in \citet{valageas2011} as:

\begin{align}
P^{(2)}(k)&=P_{13}(k)+P_{22}(k)\label{eq:valageas}\\
&=-k^2\sigma_v^2P_L(k)+\iint \frac{\mathrm{d}^3k_1\mathrm{d}^3k_2}{(2\pi)^3}\delta_D(\kk - \kk_1 - \kk_2)\frac{(\kk \cdot \kk_1)^2(\kk \cdot \kk_2)^2}{2k_1^4k_2^4}P_L(k_1)P_L(k_2)\ ,\notag\\
\sigma_v^2&=\frac{1}{6\pi^2}\int_0^{\infty}\! P_L(w)\mathrm{d}w\ . \notag
\end{align}

We group Equation (\ref{eq:nlcf}) into two terms, in order to clearly relate it to $P_{13}(k)$ and $P_{22}(k)$:
\begin{align}
\xi^{(2)}(r)&=\xi_{(13)}(r)+\xi_{(22)}(r)\ ,
\end{align}
where $\xi_{(13)}(r)$ is defined as the first term in Equation (\ref{eq:nlcf}), and $\xi_{(22)}(r)$ contains the remaining terms. The term $\xi_{(13)}(r)$ straightforwardly Fourier transforms to $P_{13}(k)$:
\begin{align}
\mathcal{F}\left\{ - \frac{1}{3}\xi_0^{-2}(0)\xi_0^{2}(r)\right\}&=-\frac{(4\pi)^2}{3(2\pi)^3}\xi_0^{-2}(0)\int_0^{\infty}\!\int_0^{\infty}\! P_L(k')k'^4j_0(k' r) j_0(kr)r^2\mathrm{d}k'\mathrm{d}r\\
&=-\frac{2}{3\pi}\xi_0^{-2}(0)\int_0^{\infty} P_L(k')k'^4\frac{\pi}{2k'^2}\delta_D(k - k')\mathrm{d}k'\notag\\
&=-\frac{1}{3}\xi_0^{-2}(0)k^2P_L(k)\ .\notag
\end{align}
This is equal to $P_{13}(k)$ from Equation (\ref{eq:valageas}):
\begin{align}
P_{13}(k)&=-k^2\sigma_v^2P_L(k) =-\frac{1}{3}\xi_0^{-2}(0)k^2P_L(k)\ .
\end{align}

The Fourier transform of  $\xi_{(22)}(r)$ contains integrals of three spherical Bessel functions. In general:
\begin{align}
\mathcal {F}\{\xi_n^{m_1}(r)\xi_n^{m_2}(r)\}&=\frac{1}{\pi^3}\int_0^{\infty}\!\int_0^{\infty}\! P_L(k_1)P_L(k_2) k_1^{m_1+2} k_2^{m_2+2}\mathrm{d} k_1 \mathrm{d} k_2 \left (\int_0^{\infty}\! j_n(k_1 r) j_{n} (k_2 r) j_0(k r) r^2 \mathrm{d}r\right ) \ .
\end{align}
We require the expressions for integrals of the type
\begin{align}
\int_0^{\infty}\! j_0(kr)j_n(k_1 r)j_n(k_2 r)r^2 \mathrm{d}r \text{,}
\end{align}
which can be found in \citet{rami1991}.

For example, to compute $\mathcal{F}\{\xi_2^0(r)^2\}$:
\begin{align}
\int_0^{\infty}\! j_2(k_1 r)& j_2 (k_2 r) j_0(k r) r^2 \mathrm{d}r = \frac{\pi\beta(k_1, k_2, k)}{32 k k_1^3 k_2^3}\left( 3(k_2^2 + k_1^2 - k^2)^2 - 4k_1^2k_2^2\right)\ ,\\ 
\beta(k_1, k_2, k)&=\begin{cases} 1 & |k_1-k_2|<k<k_1+k_2
\\
\frac{1}{2} & |k_1-k_2|=k \textit{ or } k_1+k_2=k
\\
0 & \text{ otherwise }
\end{cases}\notag
\end{align}
This means that the integral is zero unless $\kk$, $\kk_1$, and $\kk_2$ form a triangle. In the $k_1$ $k_2$ plane, we consider the region where this triangular condition is satisfied. On the boundary of that region, the Bessel integral is discontinuous. However, since the boundary has measure zero, it does not contribute to the integral. Thus, we consider only the inside of the region of the plane where $\kk$, $\kk_1$, and $\kk_2$ form a triangle, i.e. where $\beta=1$.

In the integral, we make the substitution $k_2=|\kk - \kk_1|$, and integrate over the angle $\theta$ between $\kk$ and $\kk_1$.

\begin{align}
\mathcal {F}\{\xi_2^0(r)^2\}&=\frac{1}{64\pi^3} \int\!  \mathrm{d}^3 k_1P_L(k_1)P_L(\left |\kk-\kk_1\right |)  \frac{ 3((\kk - \kk_1)^2 + k_1^2 - k^2)^2 - 4k_1^2(\kk - \kk_1)^2}{ k_1^2 |\kk - \kk_1|^2} \notag\\
&=\frac{1}{4(2\pi)^3} \iint \! \mathrm{d}^3 k_1\mathrm{d}^3k_2 P_L(k_1)P_L(k_2)\delta_D(\kk - \kk_1 - \kk_2)  \frac{ 3( \kk_1\cdot \kk_2)^2 - 2k_1^2k_2^2}{ k_1^2 k_2^2} \ .
\end{align}
This illustrates how the products of two $\xi_n^m(r)$ in configuration space correspond to a convolution in Fourier space.

The Fourier transform of all the terms in $\xi_{(22)}(r)$ is:
\begin{align}
\mathcal{F}\{\xi_{(22)}(r)\}&=\iint\! \frac{\mathrm{d}^3k_1\mathrm{d}^3k_2}{(2\pi)^3}\delta_D(\kk - \kk_1 - \kk_2)\frac{(\kk \cdot \kk_1)^2(\kk \cdot \kk_2)^2}{2k_1^4k_2^4}P_L(k_1)P_L(k_2)\ .
\end{align}
which is exactly the $P_{22}(k)$ term from Equation (\ref{eq:valageas}). Thus, the entire expression transforms to $P^{(2)}(k)$.

This agreement validates our configuration-space approach to perturbation theory with the Zel'dovich approximation. It is also remarkable how the triangular closure of the three wave vectors is connected to the integral over three spherical Bessel functions.

\section{Comparison to Numerical Results}
\label{sec:discussion}

Our goal is to understand the effects of nonlinearities on the baryon acoustic peak in the correlation function. We have shown that we can derive an expression for the first nonlinear term in configuration space from the Zel'dovich approximation. For the real-space correlation function, we are able to check our analytical expression against the Zel'dovich power spectrum from \citet{grinsteinwise1987} and \citet{valageas2011}, which is calculated using an entirely different technique. This provides analytical validation of our configuration-space approach to perturbation theory.

However, in the future we would like to extend this calculation to redshift space and to higher orders. For these applications, the Fourier equivalents do not exist, so we would have no analytical validation of the results. Therefore, we need a numerical method that can be used to test these future results. Here, we present the results from a set of Zel'dovich simulations, and show that they validate our configuration-space result, and can therefore be used to test future analytical results in redshift space and at higher orders.

A set of 100 Zel'dovich simulations, written in Python, was run with the cosmological parameters $\Omega_{\Lambda}=0.71$, $\Omega_m=0.29$, $\Omega_b= 0.045$, $h=0.7$, and $\sigma_8(z=0)=0.89$. Each simulation was a $1$ Gpc/h box with $512^3$ particles, and a cell length of $3.9$ Mpc/h. We computed the density at $4$ different redshifts in each simulation from $z=10$ and $z=0$ using a cloud-in-cell interpolation scheme. The density was then used to compute the correlation function at each time step. The average of the correlation functions over the 100 simulations allows us to see the behavior of the nonlinear term over time.

We expect the results from the simulations to agree with the analytical expression at high redshift, although the result will be noisy due to the small value of the nonlinear contribution at early times. There will be a growing deviation from the analytical result with decreasing redshift coming from the next higher order term, proportional to $D^6$.  At $z=1$, this higher order term will be smaller than 1\% of the value at the acoustic peak. At $z=0$, however, this term will cause a deviation on the order of 5\% at the BAO scale, and thus we do not expect the analytical expression to approximate well the behavior at $z=0$.  This is similar to the results from 1-loop standard perturbation theory as compared to N-body simulations \citep{matsubaraRPT2008}, where the higher order terms have the greatest contribution at low redshift.

\begin{figure}
\begin{center}
\includegraphics[scale=0.25]{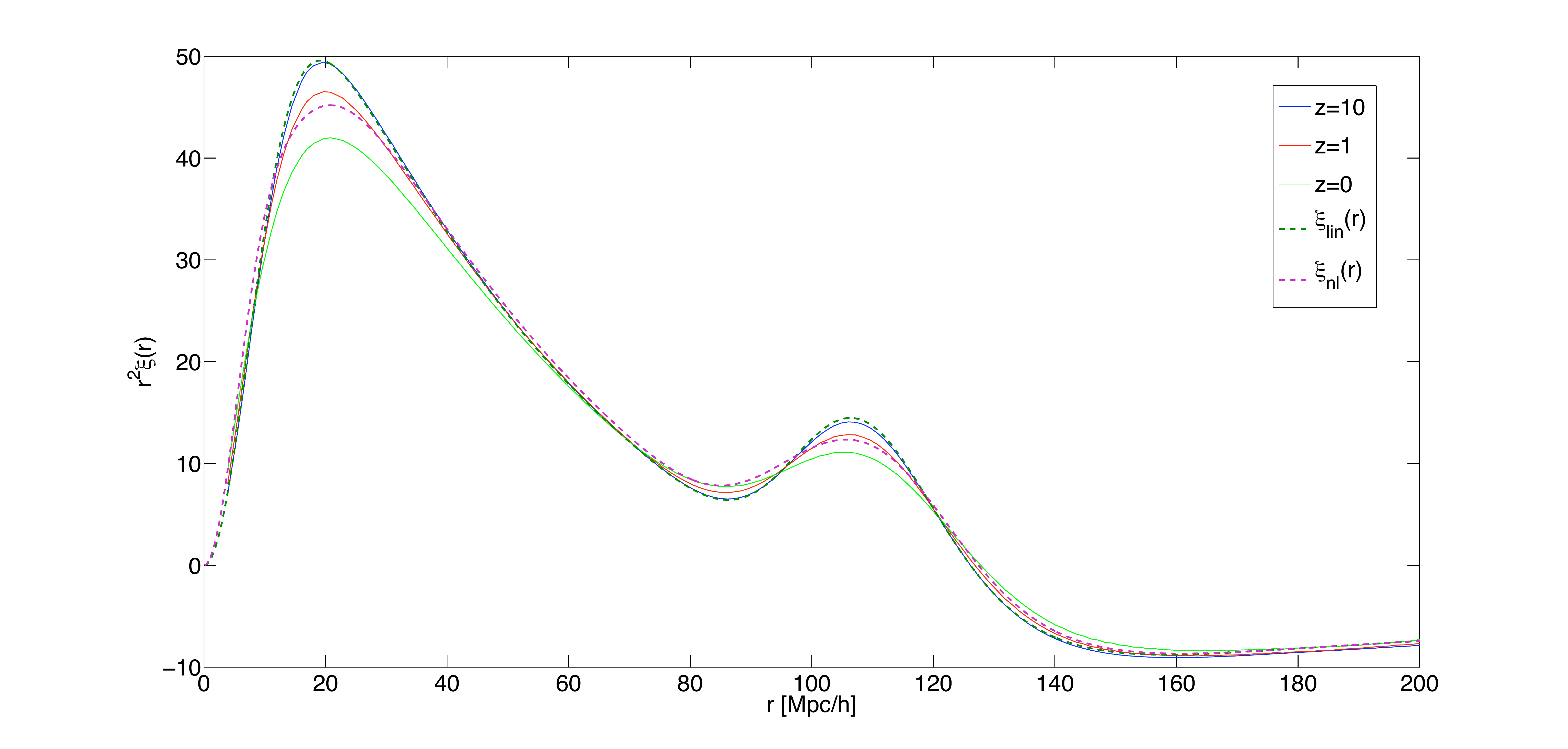}
\caption{The full correlation function. The dashed lines represent the theoretical predictions for the linear (dark green) and nonlinear (purple) correlation functions at z=0 from the Zel'dovich approximation. The solid lines are the correlation functions, normalized by $D^2$, from 100 realizations. At $z=10$, there is good agreement with linear theory. At $z=1$ the behavior of the peak is well-described by the nonlinear correlation function. At $z=0$ the first nonlinear term is not sufficient to describe the deviation from linear theory, as higher order terms become non-negligible.}
\label{fig:cfsimlin}
\end{center}
\end{figure}

Figure~\ref{fig:cfsimlin} shows the averaged correlation functions at redshifts $10$, $1$, and $0$, normalized by $D^2$ for comparison. We compare these to the theoretical predictions from the Zel'dovich approximation for the linear and nonlinear correlation functions at $z=0$. At $z=10$, the correlation function is well-approximated by linear theory. At $z=1$, the behavior of the peak is closer to that of the nonlinear correlation function. At $z=0$, higher order terms become non-negligible, so the first nonlinear term is not sufficient to fully describe the correlation function. As expected from perturbation theory, the acoustic peak is noticeably dampened at later times.

\begin{figure}
\begin{center}
\includegraphics[scale=0.25]{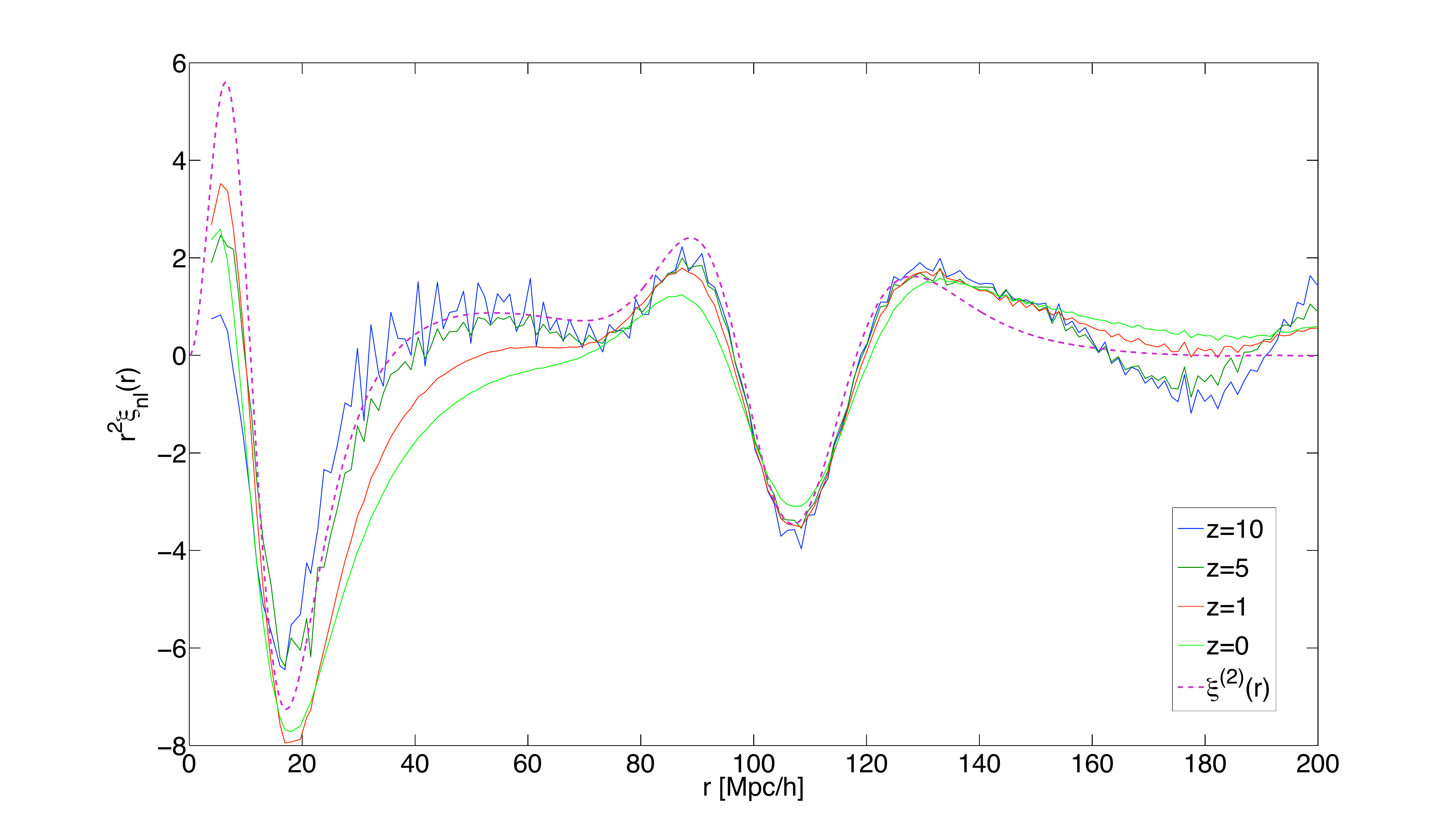}
\caption{The nonlinear correlation function. The dashed line represents the nonlinear term from the Zel'dovich approximation. The solid lines are the nonlinear term from Zel'dovich simulations, averaged over 100 realizations. At $z=10$ there is good agreement with the analytical prediction. At $z=0$, higher order terms have a non-negligible effect.}
\label{fig:cfsim}
\end{center}
\end{figure}

Figure~\ref{fig:cfsim} shows only the nonlinear contribution to the correlation function at $z=10$, $z=5$, $z=1$, and $z=0$, again normalized by $D^2$ for comparison. We also show the prediction from perturbation theory for the nonlinear term. There is overall agreement with theory at scales larger than about $20$ Mpc/h. Below this scale, we are limited by the resolution of our simulation. At scales greater than about $140$ Mpc/h, we see deviation in the nonlinear term from the theoretical prediction. We believe that this is also an artifact of the simulation, related to aliasing of the particle window function. The deviation is amplified in this figure due to the multiplication with a very large value of $r^2$. On the scale of the acoustic peak, there is good agreement between the analytical expression and the simulations. At $z=0$ we again see that the first nonlinear term is not sufficient to accurately describe the peak, because the higher order terms become non-negligible.

\newpage

\section{Conclusion}
\label{sec:conclusion}
We have shown that the first nonlinear term in the correlation function can be directly calculated in configuration space from the Zel'dovich approximation, and that this term is in agreement with both the analytical Fourier-space result and with numerical simulations. In configuration space, this nonlinear term has a simple form -- the sum of products of a broader class of linear correlation functions, $\xi_n^m(r)$. Higher order terms are in principle straightforward to compute, involving terms with products of more than two of these functions. The simple form of these higher order terms is one advantage to developing perturbation theory in configuration space.

Another advantage is that we can straightforwardly extend the real-space correlation function into redshift space. This allows us to fully characterize the nonlinear contribution to the acoustic peak in redshift space. In a future paper, we will show how this calculation, which for the Zel'dovich approximation has not been possible in Fourier space, can be done easily using our configuration-space approach.

\acknowledgements
We thank M. Neyrinck, X. Wang, and G. Lavaux for many useful discussions. This research has been supported by the Gordon and Betty Moore Foundation.

\bibliographystyle{apj}
\bibliography{zeldptbib}

\begin{thebibliography}{30}
\expandafter\ifx\csname natexlab\endcsname\relax\def\natexlab#1{#1}\fi

\bibitem[{{Bernardeau} {et~al.}(2002){Bernardeau}, {Colombi}, {Gazta{\~n}aga},
  \& {Scoccimarro}}]{bernardeau2002}
{Bernardeau}, F., {Colombi}, S., {Gazta{\~n}aga}, E., \& {Scoccimarro}, R.
  2002, \physrep, 367, 1

\bibitem[{{Blake} {et~al.}(2011){Blake}, {Kazin}, {Beutler}, {Davis},
  {Parkinson}, {Brough}, {Colless}, {Contreras}, {Couch}, {Croom}, {Croton},
  {Drinkwater}, {Forster}, {Gilbank}, {Gladders}, {Glazebrook}, {Jelliffe},
  {Jurek}, {Li}, {Madore}, {Martin}, {Pimbblet}, {Poole}, {Pracy}, {Sharp},
  {Wisnioski}, {Woods}, {Wyder}, \& {Yee}}]{blake2011}
{Blake}, C., {et~al.} 2011, \mnras, 418, 1707

\bibitem[{{Bond} \& {Efstathiou}(1984)}]{bondefstathiou1984}
{Bond}, J.~R., \& {Efstathiou}, G. 1984, \apj, 285, 45

\bibitem[{{Bouchet} {et~al.}(1995){Bouchet}, {Colombi}, {Hivon}, \&
  {Juszkiewicz}}]{bouchet1995}
{Bouchet}, F.~R., {Colombi}, S., {Hivon}, E., \& {Juszkiewicz}, R. 1995, \aap,
  296, 575

\bibitem[{{Cole} {et~al.}(2005){Cole}, {Percival}, {Peacock}, {Norberg},
  {et~al.}}]{cole2005}
{Cole}, S., {Percival}, W.~J., {Peacock}, J.~A., {Norberg}, P., {et~al.} 2005,
  \mnras, 362, 505

\bibitem[{{Crocce} \& {Scoccimarro}(2006)}]{croccescoccimarro2006}
{Crocce}, M., \& {Scoccimarro}, R. 2006, \prd, 73, 063519

\bibitem[{{Crocce} \& {Scoccimarro}(2008)}]{croccescoccimarro2008}
---. 2008, \prd, 77, 023533

\bibitem[{{de Bernardis} {et~al.}(2000){de Bernardis}, {Ade}, {Bock}, {Bond},
  {et~al.}}]{debernardis2000}
{de Bernardis}, P., {Ade}, P.~A.~R., {Bock}, J.~J., {Bond}, J.~R., {et~al.}
  2000, \nat, 404, 955

\bibitem[{{Eisenstein} {et~al.}(2005){Eisenstein}, {Zehavi}, {Hogg},
  {Scoccimarro}, {et~al.}}]{eisenstein2005}
{Eisenstein}, D.~J., {Zehavi}, I., {Hogg}, D.~W., {Scoccimarro}, R., {et~al.}
  2005, \apj, 633, 560

\bibitem[{{Grinstein} \& {Wise}(1987)}]{grinsteinwise1987}
{Grinstein}, B., \& {Wise}, M.~B. 1987, \apj, 320, 448

\bibitem[{{Hanany} {et~al.}(2000){Hanany}, {Ade}, {Balbi}, {Bock},
  {et~al.}}]{hanany2000}
{Hanany}, S., {Ade}, P., {Balbi}, A., {Bock}, J., {et~al.} 2000, \apj Letters,
  545, L5

\bibitem[{{Jain} \& {Bertschinger}(1994)}]{jainbertschinger}
{Jain}, B., \& {Bertschinger}, E. 1994, \apj, 431, 495

\bibitem[{{Kaiser}(1987)}]{kaiser1987}
{Kaiser}, N. 1987, \mnras, 227, 1

\bibitem[{{Kofman} {et~al.}(1994){Kofman}, {Bertschinger}, {Gelb}, {Nusser}, \&
  {Dekel}}]{kofman1994}
{Kofman}, L., {Bertschinger}, E., {Gelb}, J.~M., {Nusser}, A., \& {Dekel}, A.
  1994, \apj, 420, 44

\bibitem[{{Matsubara}(2008{\natexlab{a}})}]{matsubaraPTRSD2008}
{Matsubara}, T. 2008{\natexlab{a}}, \prd, 78, 083519

\bibitem[{{Matsubara}(2008{\natexlab{b}})}]{matsubaraRPT2008}
---. 2008{\natexlab{b}}, \prd, 77, 063530

\bibitem[{Mehrem {et~al.}(1991)Mehrem, Londergan, \& Macfarlane}]{rami1991}
Mehrem, R., Londergan, J.~T., \& Macfarlane, M.~H. 1991, Journal of Physics A:
  Mathematical and General, 24, 1435

\bibitem[{{Padmanabhan} {et~al.}(2007){Padmanabhan}, {Schlegel}, {Seljak},
  {Makarov}, {et~al.}}]{padmanabhan2007}
{Padmanabhan}, N., {Schlegel}, D.~J., {Seljak}, U., {Makarov}, A., {et~al.}
  2007, \mnras, 378, 852

\bibitem[{{Peebles} \& {Yu}(1970)}]{peeblesyu1970}
{Peebles}, P.~J.~E., \& {Yu}, J.~T. 1970, \apj, 162, 815

\bibitem[{{Percival} {et~al.}(2007){Percival}, {Cole}, {Eisenstein}, {Nichol},
  {Peacock}, {Pope}, \& {Szalay}}]{percival2007}
{Percival}, W.~J., {Cole}, S., {Eisenstein}, D.~J., {Nichol}, R.~C., {Peacock},
  J.~A., {Pope}, A.~C., \& {Szalay}, A.~S. 2007, \mnras, 381, 1053

\bibitem[{{Percival} {et~al.}(2010){Percival}, {Reid}, {Eisenstein}, {Bahcall},
  {et~al.}}]{percival2010}
{Percival}, W.~J., {Reid}, B.~A., {Eisenstein}, D.~J., {Bahcall}, N.~A.,
  {et~al.} 2010, \mnras, 401, 2148

\bibitem[{{Seo} \& {Eisenstein}(2003)}]{seoeisenstein2003}
{Seo}, H.-J., \& {Eisenstein}, D.~J. 2003, \apj, 598, 720

\bibitem[{{Seo} {et~al.}(2012){Seo}, {Ho}, {White}, {Cuesta}, {Ross}, {Saito},
  {Reid}, {Padmanabhan}, {Percival}, {de Putter}, {Schlegel}, {Eisenstein},
  {Xu}, {Schneider}, {Skibba}, {Verde}, {Nichol}, {Bizyaev}, {Brewington},
  {Brinkmann}, {Costa}, {Gott}, {Malanushenko}, {Malanushenko}, {Oravetz},
  {Palanque-Delabrouille}, {Pan}, {Prada}, {Ross}, {Simmons}, {Simoni},
  {Shelden}, {Snedden}, \& {Zehavi}}]{seo2012}
{Seo}, H.-J., {et~al.} 2012, ArXiv e-prints

\bibitem[{{Shandarin} \& {Zeldovich}(1989)}]{shandarin1989}
{Shandarin}, S.~F., \& {Zeldovich}, Y.~B. 1989, Reviews of Modern Physics, 61,
  185

\bibitem[{{Silk}(1968)}]{silk1968}
{Silk}, J. 1968, \apj, 151, 459

\bibitem[{{Sunyaev} \& {Zel'dovich}(1970)}]{sz1970}
{Sunyaev}, R.~A., \& {Zel'dovich}, Y.~B. 1970, APSS, 9, 368

\bibitem[{{Taruya} {et~al.}(2010){Taruya}, {Nishimichi}, \&
  {Saito}}]{taruya2010}
{Taruya}, A., {Nishimichi}, T., \& {Saito}, S. 2010, \prd, 82, 063522

\bibitem[{{Tian} {et~al.}(2011){Tian}, {Neyrinck}, {Budav{\'a}ri}, \&
  {Szalay}}]{tian2011}
{Tian}, H.~J., {Neyrinck}, M.~C., {Budav{\'a}ri}, T., \& {Szalay}, A.~S. 2011,
  \apj, 728, 34

\bibitem[{Valageas(2011)}]{valageas2011}
Valageas, P. 2011, \aap, 526, A67

\bibitem[{{Zel'dovich}(1970)}]{zeldovich1970}
{Zel'dovich}, Y.~B. 1970, Astron. Astrophys., 5, 84

\end{thebibliography}

\end{document}